\documentstyle[prl,amsfonts,twocolumn,aps,epsfig]{revtex}
\begin{document}
\title{
Spin Liquid in the Multiple-Spin Exchange model
on the Triangular lattice:  $^3$He on graphite}
\author{
G. Misguich  
\thanks{Laboratoire de Physique Th\'eorique des
  Liquides, Universit\'e P. et M. Curie, case 121, 4 Place Jussieu,
75252
  Paris Cedex. UMR 7600 of CNRS.},
B. Bernu  $^*$,
C. Lhuillier  $^*$
and C. Waldtmann
\thanks{Institut f\"ur Theoretische Physik,
Universit\"at Hannover, D-30167 Hannover, Germany}
}
\date{\today}

\maketitle

\bibliographystyle{prsty}

\begin{abstract}
Using exact diagonalizations, we investigate the $T=0$
phase diagram of the Multi-Spin Exchange (MSE) model on the
triangular lattice: we find a transition separating a
ferromagnetic phase from a non-magnetic gapped Spin
Liquid phase. Systems
far enough from the ferromagnetic transition
have a metamagnetic behavior with magnetization plateaus
at $m/m_{sat}=0$ and $1/2$.
The MSE has been proposed to describe solid
$^3$He films adsorbed onto graphite, thus we compute the MSE
heat capacity for parameters in the low density range of the 2$^{nd}$
layer and find a double-peak structure.

\end{abstract}

An increasing number of experiments on $^3$He films
have enforced the triangular lattice Multiple-Spin
Exchange (MSE) picture of the
solid second layer (\cite{gr95,sncs97} and references therein).
Following Thouless\cite{t65,rhd83}, the magnetic properties
of a 2-dimensional quantum crystal are described in spin space
by the effective Hamiltonian:
\begin{equation}
	\label{eq-ham}
	H = - \sum_P (-1)^{{\rm sign}(P)}J_P P
\end{equation}
where $P$ is any permutation operator of the spins of the lattice,
and $J_P$ equals one half of the (positive) tunneling
frequency associated to the exchange process $P$.
Whereas exchange of an even number of spins favors
antiferromagnetism, odd processes are ferromagnetic.
On the triangular lattice with spin 1/2,
two- and 3-body exchange reduce to an Heisenberg Hamiltonian
with an effective $J_2^{\rm eff}=J_2-2J_3$.
In two-dimensional $^3$He, because of stoichiometric hindrance,
the 3-body process is much more efficient than the 2-body one,
and the effective coupling constant is ferromagnetic. 
However, it was suspected since a long time that $n$-particle
terms with $n>3$  could not be ignored \cite{g90,gr95,sncs96}.
Very recently, a thorough analysis 
of susceptibility and heat capacity
measurements enabled Roger and the Grenoble group \cite{rbbcg98}
to establish the density dependence of the $J$'s and the
importance of the 4-spin exchange $J_4$ in a large density
range of the 2$^{\rm nd}$ layer: from the 2$^{\rm nd}$ layer
solidification
(AF solid) to the third layer promotion (F solid). This is in
accordance
with recent Path Integral Monte Carlo calculations 
which estimate $J_2^{\rm eff}/J_4\simeq -2\pm 1$,
$J_5/J_4\simeq 1\pm .5$ and $J_6/J_4\simeq 1 \pm .5$ \cite{bc97}.

As first suggested by Roger in 1990 \cite{r90b}, 4-spin exchange
strongly frustrates the system.
Momoi, Kubo and Niki have recently studied the $J_2^{\rm eff}$-$J_4$
model
in the classical and semi-classical (spin-wave) limits and found
numerous ordered phases: a ferromagnetic, a  4-sublattice ferrimagnetic
phase, 3- and 4-sublattice AF phases and a chiral ordered
phase \cite{mkn97,km97}. Our $SU(2)$ Schwinger-Boson
analysis of the $J_2^{\rm eff}$-$J_4$-$J_5$ Hamiltonian also pointed
to a rich
phase diagram with N\'eel as well as helicoidal phases\cite{mbl98}.

In the present work, exact diagonalizations results
show that  none of these T=0 Long Range Ordered (LRO) AF phases
survive spin-1/2 quantum fluctuations. Instead, in the AF region,
the system is a  Quantum  Spin Liquid  (QSL)
with short range spin-spin correlations,
as suggested by Ishida {\it et al.} \cite{imyf97}
and Kubo {\it et al.} \cite{km97}.

We truncate Eq. (\ref{eq-ham}) to 
the following simplest cyclic exchange patterns:
\begin{eqnarray}\label{Hmulti}
H=J^{\rm eff}_2 \sum_{
\begin{picture}(17,10)(-2,-2)
	\put (0,0) {\line (1,0) {12}}
	\put (0,0) {\circle*{5}}
	\put (12,0) {\circle*{5}}
\end{picture}
} P_{2}
+J_4 \sum_{
\begin{picture}(26,15)(-2,-2)
        \put (0,0) {\line (1,0) {12}}
        \put (6,10) {\line (1,0) {12}}
        \put (0,0) {\line (3,5) {6}}
        \put (12,0) {\line (3,5) {6}}
        \put (6,10) {\circle*{5}}
        \put (18,10) {\circle*{5}}
        \put (0,0) {\circle*{5}}
        \put (12,0) {\circle*{5}}
\end{picture}
} \left( P_{4}+P_{4}^{-1}\right) \\
-J_5 \sum_{
\begin{picture}(26,15)(-2,-2)
        \put (0,0) {\line (1,0) {24}}
        \put (6,10) {\line (1,0) {12}}
        \put (0,0) {\line (3,5) {6}}
        \put (18,10) {\line (3,-5) {6}}
        \put (6,10) {\circle*{5}}
        \put (18,10) {\circle*{5}}
        \put (0,0) {\circle*{5}}
        \put (12,0) {\circle*{5}}
        \put (24,0) {\circle*{5}}
\end{picture}
} \left( P_{5}+P_{5}^{-1}\right) \nonumber
+J_6 \sum_{
\begin{picture}(26,30)(-2,-15)
        \put (6,10) {\line (1,0) {12}}
        \put (6,-10) {\line (1,0) {12}}
        \put (0,0) {\line (3,5) {6}}
        \put (0,0) {\line (3,-5) {6}}
        \put (18,10) {\line (3,-5) {6}}
        \put (18,-10) {\line (3,5) {6}}
        \put (6,10) {\circle*{5}}
        \put (6,-10) {\circle*{5}}
        \put (18,10) {\circle*{5}}
        \put (18,-10) {\circle*{5}}
        \put (0,0) {\circle*{5}}
        \put (12,0) {\circle*{2}}
        \put (24,0) {\circle*{5}}
\end{picture}
} \left( P_{6}+P_{6}^{-1}\right) \nonumber
\end{eqnarray}
The spin-1/2 permutation operators can be rewritten with usual spin
operators (Pauli matrices):
$ P_{ij} = 2 {\bf S}_{i}.{\bf S}_{j} +1/2 $.  $ P_{1234} +h.c.$
and $ P_{12345} +h.c.$
(resp.  $ P_{123456} +h.c.$) are  polynoms of degree two 
(resp. three) in $\bf{S}_{i}.\bf{S}_{j}$ \cite{mbl98,go88}.
The quantum phase diagram of this model is studied through
the analysis of the finite size scaling of the low energy
spectra of samples
subjected to various boundary conditions
(twisted or periodic with different shapes).

\begin{figure}
\begin{center}
\mbox{\psfig{figure=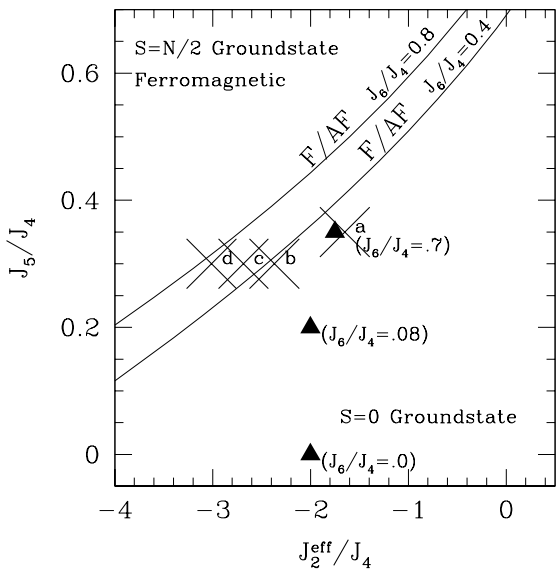,width=8cm}}
\caption[99]{Phase diagram for $J_2^{\rm eff}<0$.
The  solid line is the $T=0$ transition line 
between ferromagnetic and antiferromagnetic phases.
The crosses are Roger {\it et al}'s results\cite{rbbcg98}
for four 2$^{\rm nd}$ layer
densities ($nm^{-2}$): $\rho_2^a=6.5$ ($J_6/J_4\simeq0.7$),
$\rho_2^b=7.0$, $\rho_2^c=7.65$ and $\rho_2^d=7.8$.
For $b$, $c$ and $d$ $J_6/J_4\simeq0.4$
(the size of the crosses may underestimate
the uncertainties in the parameters).
Two of the black triangles indicate the sets of coupling parameters
corresponding to Fig. \ref{Egap},\ref{cv},\ref{magn}.
The upper one is the point close to the frontier mentioned in the text.}
\label{diag}
\end{center}
\end{figure}

{\it Ferro-antiferro transition}.---
We first look at the
$T=0$ ferro-antiferro
transition line of Eq. (\ref{Hmulti}). This line
(see Fig. \ref{diag}) is
determined from about 50 spectra ($N=19$) in the $J_n$ space,
with most points
near the transition. Far all cases, the ground-state is either
an $S=0$ or an
$S=N/2$ state. This excludes the possibility of a {\it uuud} phase
found in the classical calculations of Kubo {\it et al.} \cite{km97}, which
is ferrimagnetic and has a total spin $S=N/4$.
The line where $J_{\chi}$, the $1/T^2$ coefficient of the
susceptibility,
vanishes stands roughly parallel to the $T=0$ F/AF line,
inside the AF region.
The density dependence of the $J_n$'s proposed by Roger
{\it et al.} \cite{rbbcg98} for the second layer
(see crosses in Figure \ref{diag}) leads us to conclude
that a T=0 transition to ferromagnetism occurs at
$\rho_2\simeq 6.8 \pm 0.3 nm^{-2}$
whereas $J{\chi}$ is zero at
$\rho_2 \simeq 6.5 nm^{-2}$ \cite{rbbcg98}.

{\it No Long Range Order}.---
The nature of the non magnetic or antiferromagnetic (AF)
phase is a more challenging question.
We first look for signatures of
N\'eel Long Range Order (NLRO). NLRO is characterized
on finite systems by very stringent spectral properties
\cite{blp92}:
{\it i}) The symmetry breakings associated to the order parameter 
are embodied in a family of $\sim N^\alpha$ low lying levels
collapsing to the ground-state in the thermodynamic limit
($\alpha$ is the number of magnetic sublattices).
These levels with definite space and $SU(2)$ symmetries
and dynamical properties
should appear directly below the first magnon excitations. {\it ii})
The finite-size scalings of these levels are known.
In particular,
the ground-state energy per site $E(S=0,N)/N$ has corrections scaling
as $N^{-3/2}$,
the $\Delta S=1$ spin gap $E(S=1,N)-E(S=0,N)$ goes to zero as $N^{-1}$.
None of these prescriptions is obeyed by the MSE spectra
in the ``S=0 ground-state'' region displayed Fig. \ref{diag},
whatever the twisted or shape boundary conditions may be.
Thus, we exclude any commensurate or non-commensurate NLRO.

\begin{figure}
\begin{center}
\mbox{\psfig{figure=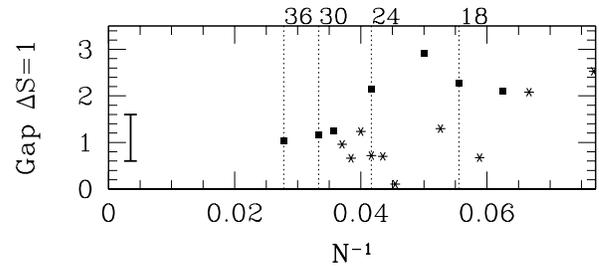,width=8cm}}
\caption[99]{
	Spin gap plotted as a function of 1/N
	for $J_2^{\rm eff}=-2$, $J_4=1$. The
	vertical bar is the $N=\infty$ extrapolation
	made out of the $E/N$ $\leftrightarrow$ $\Delta$
	correlation analysis.
	}
\label{Egap}
\end{center}
\end{figure}

{\it A $J_2^{eff} - J_4$ model}.---
We have analyzed in detail the $J_2^{\rm eff}=-2,J_4=1$ model
for nearly all possible systems from $N=6$ to 30 and for $N=36$.
We noticed two different
scaling behaviors: samples with $N$ multiple of 4 or 6 have a low
ground-state energy increasing with $N$ whereas others samples have 
a high ground-state energy decreasing with N. The energies
of both families merge for $N_0\simeq 40$. Our interpretation
is the following:
$N_0$ is a crossover size above which the system is not
anymore sensitive
to boundary conditions and this is the signature
of a finite length scale $\xi=\sqrt{N_0}$ in the ground-state
wave function. This is supported by two facts: a fast decay 
of the spin-spin correlations with
distance and a non-vanishing
spin gap $\Delta$ in the thermodynamic limit. $\Delta S=1$ gaps are
plotted in Figure \ref{Egap}: the two families of samples
(squares for
$N$ multiple of 4 or 6 and stars for others)
have gaps of the order of 1 for $N>24$.  An
estimation of the $N=\infty$ gap is possible using
the strong correlation between $E/N$ and $\Delta$, the
result is $\Delta \simeq 1.1\pm 0.5$
(details will be given elsewhere).

These data point to a Quantum Spin Liquid state with a gap 
of the order of 1 for $J_2^{\rm eff}=-2$ and $J_4=1$.
This gap and the sensitivity to the geometrical shapes and
boundary conditions of the small samples 
suggest a valence-bond picture of the ground-state and of the
first triplet excitation.

Because of the strong frustration between the
effective first neighbor ferromagnetic Heisenberg term
($J_2^{\rm eff}$)
and the 4-spin antiferromagnetic exchange,
it is the triangular 6-spin plaquette
which is the first system with a paramagnetic ($S=0$) ground-state
and a significant gap\footnote{
	For $N=24$ (resp. $N=12$) one can compare the spectrum
	of the sample built with four (resp. two) 6-site triangles
	with the spectra computed from other shapes.
	The triangle-compatible shape gives the lowest energy
	and the largest gap
	(others have energies and gaps comparable with the
	frustrated family ones).}.
However the ground-state wave function
is not a naive tensorial product of $S=0$ independent
triangle wave functions:
such an approximation gives a very high ground-state energy
($-2.8$ to be compared to $-4$) and largely underestimates the gap
(.4 to be compared to 1). This quasi independent triangle picture
is thus deeply renormalized by resonances.

Analysis of the low lying levels in the $S=0$ subspace leads to
the same conclusion: the number of singlets below the first
triplet level is very small ($\le 10$ all degeneracy taken into account).
As a comparison, this is very different
from the Kagom\'e case where
a continuum of singlets state is found in the magnetic gap
\cite{lblps97,web98}. Thus, this system seems a very example of a
short range Resonating Valence Bond state with a clear cut gap in a
translationally invariant 2-dimensional spin-1/2 model. 
Consequently, both the low temperature
specific heat and spin susceptibility are thermally activated
($\chi(T)\sim e^{-\Delta/T}$, $C_V(T)\sim e^{-\Delta/T}$).
However the gap decreases rapidly when approaching
the AF/F transition and we are not yet able to decide if the gap
vanishes at the transition to ferromagnetism
or before. Indeed, experimental results of
Ishida and collaborators  seem to indicate either a very small
gap or a quantum critical behavior\cite{imyf97}.

\begin{figure}
\begin{center}
\mbox{\psfig{figure=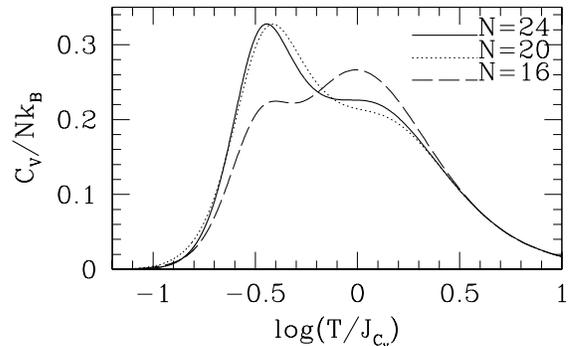,width=8cm}}
\caption[99]{
Heat capacity versus temperature for
three sizes. Coupling parameters are
$J_2^{\rm eff}/J_4=-2$, $J_5/J_4=0.2$, $J_6/J_4=0.08$.
For these values $J_{C_v}= 0.93J_4$ and the $N=\infty$ spin gap is 
of the order of $J_4/2$. Due to finite-size effects, on $N=20$ and $24$ the
high temperature peak ($T\simeq J_{C_V}$) only shows
up as a shoulder.}
\label{cv}
\end{center}
\end{figure}

{\it 2$^{\rm nd}$ layer heat capacity}.---
We compute the MSE heat capacity for $J_2^{\rm eff}/J_4=-2$,
$J_5/J_4=0.2$, $J_6/J_4=0.08$ with $N=16,20$ and $24$
(Fig. \ref{cv}) and find
a clear low temperature peak.
The entropy at $T/J_{C_v}=.5$ is $\simeq 0.4 N ln(2)$
and the low temperature peak is thus likely to remain in the
thermodynamic limit. 
It also subsists for a relatively large range 
of competing $J_n$. Such a low temperature peak is
characteristic of different highly
frustrated systems\cite{e89,r90b}.
The high temperature peak is located
at a temperature of the order of $J_{C_v}$ ($J_{C_v}$ is the
leading coefficient of the $1/T$ expansion of the specific heat:
$C_v = 9/4 (J_{C_v}/T)^2+{\mathcal O}(1/T^3)$,
its expression as a function of the $J$'s
is given in \cite{r97a}).
The low temperature peak height and location
do not only depend on $J_{C_v}$
but also on the relative values of the coupling parameters.
A better agreement between present results and experimental results
\cite{imyf97} is expected by a fine tuning
of the coupling parameters. Indeed, moving towards the boundary line
between AF and F phases both decreases the gap and shifts
the low temperature peak towards lower temperature.

{\it Low energy degrees of freedom}.---
To understand the excitations responsible for this low energy
peak we looked for possible common
properties of low lying levels: it appears that most
of these levels have a significant projection on the subspace
engendered
by spin-1 diamond tilings.  This subspace $\mathcal E$ is
defined by the
non-orthogonal family of wave functions:
\begin{equation}
	\left|\Psi\right>=
	\bigotimes_{d=1,\cdots,\frac{N}{4}}
	\left|S_d=1,S_d^z\in\{-1,0,1\}\right>
\end{equation}
where $\left|S_d=1,S_d^z\right>$ is the 
$S=1,S^z=S_d^z$ state, symmetric with respect to the small
diagonal of the $d^{th}$ diamond.
For $N=16$ the projections of the exact low lying levels
on $\mathcal E$ range from $10\%$ to $50\%$
for nearly all the states involved in the peak (and drop
under $1\%$ for higher energy states).
These numbers are very large compared
with the expectation values of the projection 
of a random $S=0$ or $S=4$ wave function, which  are,
for the same lattice size, of the order of $1\%$. 
Since each tiling of the lattice gives $3^\frac{N}{4}$
independent states, the entropy associated with $\mathcal E$
is a least
$ln(3^{\frac{N}{4}})\simeq 0.4 N ln(2)$, in agreement with the low
temperature peak entropy found in our samples.

These results lead to the following picture:
At high temperature down to $T\simeq J_{C_v}$ the degrees of freedom
are essentially random spin-1/2.
For $T$ less than $J_{C_v}$,
the thermal wavelength increases and near neighbor spins behave
coherently as weakly-ferromagnetic entities (pseudo spin-1).
This explains the low temperature peak.
At $T\simeq \Delta$ (spin gap), the 4-spin exchange
coupling creates larger clusters and the system
organizes itself as a QSL.

\begin{figure}
\begin{center}
\mbox{\psfig{figure=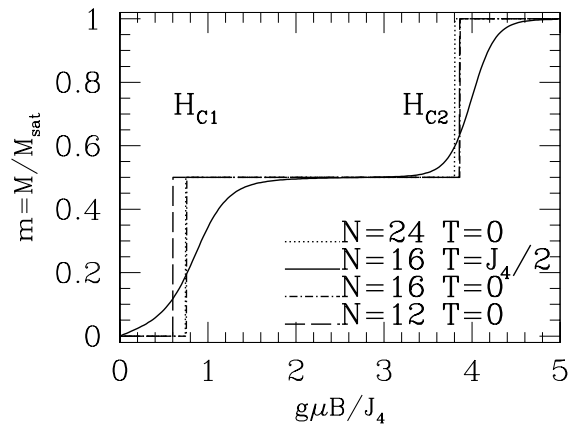,width=8cm}}
\caption[99]{Magnetization versus magnetic field $B$. Exchange parameters
are those of Fig. \ref{cv}. Since in
$^{3}$He, $J_4\simeq 2mK$ and $g\mu=1.5mK.T^{-1}$,
$H_{C1}$ is in the Telsa range.}
\label{magn}
\end{center}
\end{figure}

{\it Magnetization}.---
Now we apply a magnetic field and look for the spin $S$ of
ground-state
(T=0 magnetization).
The large gap between the sectors of total spins $S=S_{max}/2$
and $S=S_{max}/2+1$ indicates that exciting one
diamond to its $S=2$ state costs a large energy.
This feature gives rise to a low temperature 
plateau\footnote{
	Related
	phenomena have already been encountered in the
	Heisenberg model with Ising anisotropy \cite{nm86}, in
	the  MSE model on the square lattice \cite{rh90},
	on Heisenberg ladders or chains \cite{chp97,t98}.
	An $m=0$ plateau has been observed in the spin-ladder
	compound $Cu_2(C_5H_{12}N_2)_2Cl_4$ \cite{cclpmm97,hpl96}.
} at magnetization $m=1/2$ (Fig. \ref{cv})
which has also been found in the
classical variational picture by Kubo {\it et al}.\cite{km97}.
For the same coupling parameters as Fig. \ref{cv}, the figure
\ref{magn} shows two metamagnetic
transitions at $H_{C1}$ and $H_{C2}$ and the magnetization is
completely quantized\footnote{
	This, as well as the symmetries and degeneracies of the
	low lying levels of the spectra, will be discussed elsewhere
	in relation with possible extension of the Lieb-Schultz-Mattis
	theorem to 2-dimensional magnets\cite{oya97,a88}.}
: $m=0,1/2$ or $1$. Thanks to
the large gaps ($S=0 \rightarrow 1$ and $S=N/4 \rightarrow N/4+1)$
these plateaus survive thermal excitations.
Close to the ferromagnetic transition (upper black triangle of
Fig. \ref{diag}) both gaps decrease and the overall shape of the
energy versus magnetization anticipates a
transition to a phase where the magnetization curve is strongly
non linear and reminiscent of the metamagnetic
transitions
(see\cite{rh90} about metamagnetism in 3D solid $^3$He).

In this letter we have shown that the spin-1/2 MSE model, in the
range of parameters relevant for the description of $^3$He on
graphite,
exhibits a transition from an antiferromagnetic phase at low
density to a ferromagnetic
one at high density. The AF phase has no N\'eel Long Range Order and
is a Quantum Spin Liquid phase.
In this phase the specific heat has two peaks.
The lowest one testifies the building of a  QSL
out of short range pseudo spin-1. This picture is
consistent with magnetization and heat capacity measurements.
The microscopic arrangement of the spins may
also show up as plateaus in the magnetization
and $T=0$ metamagnetic transitions between zero,
half polarized and fully polarized phases.

Acknowledgments:
We have benefited from very interesting
discussions with C. B\"auerle, H. Fukuyama, H. Godfrin,
 K. Kubo, M. Roger and  J. Saunders.
Computations were performed on CRAY C94,C98 and T3E-256 at the
Institut de D\'eveloppement des Recherches en Informatique
Scientifique
of C.N.R.S. under contracts 960076/964091 and on 
CRAY T3E-512 of the Zentralinstitut f\"ur 
Angewandte Mathematik, Forschungszentrum J\"ulich.

\end{document}